\documentclass[10pt,conference, a4paper]{IEEEtran}
\usepackage{bbm}
\usepackage{cite}
\usepackage{amsfonts}
\usepackage{mathrsfs}
\usepackage{graphicx}
\usepackage{amssymb}
\usepackage{latexsym}
\usepackage{amsmath}
\usepackage{stfloats}
\usepackage{cases}
\usepackage{setspace}
\usepackage{bigstrut}
\usepackage{bm}
\usepackage{url}
\usepackage{color}
\usepackage{caption}
\usepackage{subfigure}
\usepackage{booktabs}
\usepackage[ruled]{algorithm2e}
\usepackage[colorlinks,linkcolor=black,citecolor=black,anchorcolor=black,urlcolor=black]{hyperref}

\begin{document}

\title{TurboNet: A Model-driven DNN Decoder Based on Max-Log-MAP Algorithm for Turbo Code}
\author{Yunfeng~He$^\ast$, Jing~Zhang$^\ast$, Chao-Kai~Wen$^\dagger$, and Shi~Jin$^\ast$\\
$^\ast$National Mobile Communications Research Laboratory, Southeast University, Nanjing, China\\
$^\dagger$Institute of Communications Engineering, National Sun Yat-sen University, Kaohsiung 804, Taiwan\\
Email: \{heyunfeng, jingzhang, jinshi\}@seu.edu.cn$^\ast$, chaokai.wen@mail.nsysu.edu.tw$^\dagger$\\
}
\maketitle

\begin{abstract}
This paper presents TurboNet, a novel model-driven deep learning (DL) architecture for turbo decoding that combines DL with the traditional max-log-maximum \emph{a posteriori} (MAP) algorithm. To design TurboNet, we unfold the original iterative structure for turbo decoding and replace each iteration by a deep neural network (DNN) decoding unit. In particular, the DNN decoding unit is obtained by parameterizing the max-log-MAP algorithm rather than replace the whole decoder with a black box fully connected DNN architecture. With the proposed architecture, the parameters can be efficiently learned  from training data, and thus TurboNet learns to appropriately use systematic and parity information to offer higher error correction capabilities and decrease computational complexity compared with existing methods. Furthermore, simulation results prove TurboNet’s superiority in signal-to-noise ratio generalizations.

\end{abstract}

\section{Introduction}\label{Sec:Introduction}
Recently, deep learning (DL) has made remarkable achievements in the fields of computer vision and natural language processing and it has been adopted for application in channel decoding. The data-driven DL approach in~\cite{IEEEondeep:Gruber} converted the decoding task into the pure idea of \emph{learning to decode} by optimizing the general black box fully connected deep neural network (FC-DNN). Despite the advantage of one-shot decoding (i.e., no iterations), the FC-DNN based decoder is short of expert knowledge, which in turn renders the FC-DNN decoder unaccountable and fundamentally restricted by its dimensionality. Training any neural network in practice is impossible because the training complexity increases exponentially along with block length (e.g., for a Turbo code with length of $K=40$, ${{2}^{40}}$ different codewords exist)~\cite{IEEEanartificial:Wang}. In addition, a recurrent neural network (RNN) architecture containing two layers of bidirectional gated recurrent units was adopted to learn the BCJR algorithm~\cite{IEEEcommunicationalgorithms:Kim}. The aforementioned data-driven decoding methods count on a large amount of data to train numerous parameters, thereby converging slowly and suffering a high computational complexity.

To address the aforementioned issues, the model-driven DL approach can be used instead. The concept of a ``soft'' Tanner graph was proposed in~\cite{IEEElearningto:Nachmani}, where weights were assigned to the Tanner graph of the belief propagation (BP) algorithm to obtain a deep neural network (DNN). These weights were learned to properly weight messages transmitted in Tanner graph, thereby improving the performance of BP algorithm. A large number of multiplications were required in~\cite{IEEElearningto:Nachmani}; thus, the authors in~\cite{IEEEneuraloffset:Lugosch} proposed a min-sum algorithm with trainable offset parameters to reduce the computational complexity of the algorithm. The aforementioned DNN-based BP decoder was transformed into an RNN architecture in~\cite{IEEEdeeplearning:Nachmani} named BP-RNN decoder by unifying the weights in each iteration, thereby reducing the number of parameters without sacrificing the performance. In addition, a trainable relaxation factor was introduced to improve the performance of this BP-RNN decoder.

In sum, two central limitations are inherent in current DL-based decoding methods. First, existing data-driven approaches rely on vast training parameters. Second, the aforementioned model-driven decoding algorithms are all based on the BP algorithm; however, whether these algorithms could be applied to sequential codes (e.g., Turbo code) to improve the performance remains unknown. To address the limitations, this paper presents TurboNet, a novel model-driven DL architecture for turbo decoding that combines DL with the traditional max-log-maximum \emph{a posteriori} (MAP) algorithm. TurboNet is constructed based on the domain knowledge in turbo decoding algorithms and employing several learnable parameters. More specifically, the original iterative structure is unfolded to obtain an ``unrolled'' (i.e., each iteration is considered separately) structure, and the max-log-MAP algorithm is parameterized. With the design, the parameters can be determined via training data more efficiently than the existing black box FC-DNN~\cite{IEEEanartificial:Wang} and RNN~\cite{IEEEcommunicationalgorithms:Kim} architectures. Our TurboNet decoder exhibits better performance compared with the traditional max-log-MAP algorithm for turbo decoding with different code rates (i.e., $1/2$ and $1/3$) and contains considerably fewer parameters compared with the neural BCJR decoder proposed in~\cite{IEEEcommunicationalgorithms:Kim}. Furthermore, the proposed TurboNet decoder shows strong generalizations; that is, TurboNet is trained at a special signal-to-noise ratio (SNR) and outperforms the max-log-MAP algorithm at a wide range of SNRs.

\section{TurboNet}
\label{section3}
To obtain the model-driven DL architecture for turbo decoding, we briefly describe the system model in Section~\ref{subsection:system model} and the traditional max-log-MAP algorithm in Section~\ref{subsction:the max-log-map algorithm}. The architecture and details of TurboNet are elaborated in Section~\ref{subsection:dnn turbo decoder}, including a redefined function that evaluates network loss.
\subsection{System Model}
\label{subsection:system model}
At the transmitter, a binary information sequence $\mathbf{u}$ is encoded by a turbo encoder that contains two identical recursive systematic convolutional encoders (RSCEs). The generator matrix of the RSCE is ${[1,~{{{g}_{1}} ( D)}/{{{g}_{0}} (D)}]}$, where ${{{g}_{0}}(D)=1+{{D}^{2}}+{{D}^{3}}}$ and ${{{g}_{1}}(D)=1+D+{{D}^{3}}}$~\cite{IEEE3GPP}. The feedthrough passes one block of $K$ information bits ${{u}_{k}}$, $k=0,~1,~\ldots,~K-1$, to the output of the encoder, which are then referred to as systematic bits $x_{k}^{s}={{u}_{k}}$. The first RSCE generates a sequence of parity bits $x_{k}^{1p}$ from the systematic bits, and the second RSCE generates a sequence of parity bits $x_{k}^{2p}$ from $\tilde{\mathbf{u}}$, which is an interleaved sequence of the systematic bits. ${{S}_{R}=\{0,~1,~\dots,~7\}}$ is the set of all ${{2}^{3}}$ RSCE states. The codeword $\{ x_{k}^{s},x_{k}^{1p},x_{k}^{2p}\}$ consisting of $N=3K$ bits is then modulated and transmitted over an additive white Gaussian noise (AWGN) channel. At the receiver, a soft-output detector computes reliability information in the form of log-likelihood ratios (LLRs) for the transmitted bits. The resulting LLRs $\{ y_{k}^{s},y_{k}^{1p},y_{k}^{2p}\}$ indicate the probability of the corresponding bits being a binary 1 or 0. 

\subsection{Max-Log-MAP Algorithm}
\label{subsction:the max-log-map algorithm}
A traditional turbo decoder introduced in~\cite{IEEEturbocode:Berrou} contains two soft-input soft-output (SISO) decoders, which have the same structure. Therefore, we only introduce decoder 1 in detail as follows. The MAP algorithm is used in decoder 1 to compute \emph{a posteriori} LLR for information bit as shown as follows:
\begin{equation}
\label{Eq:posteriori llr}
\begin{aligned}
L\left( \left. {{u}_{k}} \right|\mathbf{y} \right)=\log \left\{ \frac{P\left( \left. {{u}_{k}}=1 \right|\mathbf{y} \right)}{P\left( \left. {{u}_{k}}=0 \right|\mathbf{y} \right)} \right\}=\log \left\{ \frac{\sum\limits_{{{S}^{1}}}{P\left( {s}',s,\mathbf{y} \right)}}{\sum\limits_{{{S}^{0}}}{P\left( {s}',s,\mathbf{y} \right)}} \right\},
\end{aligned}
\end{equation}
where ${s}'$ and $s$ represent the states of the encoder at time $k-1$ and $k$, respectively; and the sequence ${\mathbf{y}=\{ {{y}_{k}}\}=\{ y_{k}^{s},y_{k}^{1p}\}}$ is made up of the LLRs of systematic bits and corresponding parity bits. ${{S}^{1}}$ is the set of ordered pairs $({s}',s)$ corresponding to all state transitions ${s}'\to s$ caused by data input ${{u}_{k}}=1$ and ${{S}^{0}}$ is similarly defined for ${{u}_{k}}=0$. All of the aforementioned state transitions are given in detail in the Table~\ref{tab:state transitions}.
\begin{table}[htb]
	\caption{State transitions of RSCE}
	\label{tab:state transitions}
	\centering
	\begin{tabular}{ccccccccc}
		\toprule
		${s}'$& 0 & 1 & 2 & 3 & 4 & 5 & 6 & 7  \\
		\hline
		${s:{{u}_{k}=0}}$ & 0 & 4 & 5 & 1 & 2 & 6 & 7 & 3\\
		${s:{{u}_{k}=1}}$ & 4 & 0 & 1 & 5 & 6 & 2 & 3 & 7\\
		\bottomrule
	\end{tabular}
\end{table}

On the basis of the Bayes formula, we obtain
\begin{equation}
\label{Eq:p(s',s,y)}
\begin{aligned}
P\left( s',s,\mathbf{y} \right)&=P\left( s',s,{{\mathbf{y}}_{j<k}},{{y}_{k}},{{\mathbf{y}}_{j>k}} \right) \\ 
&=P\left( \left. {{\mathbf{y}}_{j>k}} \right|s \right)P\left( \left. {{y}_{k}},s \right|s' \right)P\left( s',{{\mathbf{y}}_{j<k}} \right) \\ 
&={{\beta }_{k}}\left( s \right){{\gamma }_{k}}\left( s',s \right){{\alpha }_{k-1}}\left( s' \right),  
\end{aligned}
\end{equation}
where ${{\alpha }_{k-1}}\left( s' \right)=P\left( s',{{\mathbf{y}}_{j<k}} \right)$ and ${{\beta }_{k}}\left( s \right)=P\left( \left. {{\mathbf{y}}_{j>k}} \right|s \right)$ can be computed through the forward and backward recursions~\cite{IEEEoptimaldecoding:Bahl}:
\begin{equation}
\label{Eq:alpha}
\begin{aligned}
{{\alpha }_{k}}\left( s \right)=\sum\limits_{s'\in {{S}_{R}}}{{{\alpha }_{k-1}}\left( s' \right){{\gamma }_{k}}\left( s',s \right)}
\end{aligned}
\end{equation}
\begin{equation}
\label{Eq:beta}
\begin{aligned}
{{\beta }_{k-1}}\left( s' \right)=\sum\limits_{s\in {{S}_{R}}}{{{\beta }_{k}}\left( s \right){{\gamma }_{k}}\left( s',s \right)}
\end{aligned}
\end{equation}
with initial conditions ${{{\alpha }_{0}}(0)=1}$, ${{{\alpha }_{0}}(n)=0}$ for $n\ne 0$, and ${{{\beta }_{K}} (0)=1}$, ${{{\beta}_{K}}(n)=0}$ for $n\ne 0$. Moreover, ${{\gamma}_{k}}(s',s)=P({{y}_{k}},s|s')$ is computed as follows~\cite{IEEEiterative:Bauch}:
\begin{equation}
\label{Eq:gamma2}
\begin{aligned}
{{\gamma }_{k}}\left( {s}',s \right)=\exp \left\{ \frac{1}{2}\left( x_{k}^{s}y_{k}^{s}+x_{k}^{1p}y_{k}^{1p} \right)+\frac{1}{2}{{u}_{k}}L\left( {{u}_{k}} \right) \right\},
\end{aligned}
\end{equation}
where $L\left( {{u}_{k}} \right)$ is the \emph{a priori} probability LLR for the bit ${{u}_{k}}$. Given that $L(\left. {{u}_{k}} \right|\mathbf{y})$ is the sum of the systematic bit LLR $y_{k}^{s}$, the \emph{a priori} probability LLR $L({{u}_{k}})$, and the extrinsic LLR ${{L}_{e}}\left( {{u}_{k}} \right)$, we obtain
\begin{equation}
\label{Eq:le}
\begin{aligned}
{{L}_{e}}\left( {{u}_{k}} \right)=L(\left. {{u}_{k}} \right|\mathbf{y})-y_{k}^{s}-L({{u}_{k}}),
\end{aligned}
\end{equation}
which can be used as the \emph{a priori} probability LLR input of the subsequent SISO decoder 2 after it is interleaved.

The log-MAP algorithm introduced in~\cite{IEEEOptimal:Robertson} evaluates ${{\alpha }_{k-1}}\left( s' \right)$ and ${{\beta }_{k}}\left( s \right)$ in logarithmic terms using the Jacobian logarithmic function ${{\max }^{*}}\left( x,y \right)=\max \left( x,y \right)+\log \left( 1+{{e}^{-\left| x-y \right|}} \right)$, as shown as follows:
\begin{equation}
\label{Eq:log_alpha}
\begin{aligned}
{{\bar{\alpha }}_{k}}\left( s \right)=\underset{{s}'\in {{S}_{R}}}{\mathop{\max }}\,{{}^{*}}\left( {{{\bar{\alpha }}}_{k-1}}\left( {{s}'} \right)+{{{\bar{\gamma }}}_{k}}\left( {s}',s \right) \right)
\end{aligned}
\end{equation}
and
\begin{equation}
\label{Eq:log_beta}
\begin{aligned}
{{\bar{\beta }}_{k-1}}\left( {{s}'} \right)=\underset{s\in {{S}_{R}}}{\mathop{{{\max }^{*}}}}\,\left( {{{\bar{\beta }}}_{k}}\left( s \right)+{{{\bar{\gamma }}}_{k}}\left({s}',s\right)\right),
\end{aligned}
\end{equation}
where ${{\bar{\alpha }}_{k}}\left( s \right)$, ${{\bar{\beta }}_{k}}\left( s \right)$, and ${{\bar{\gamma }}_{k}}\left( s',s \right)$ represent the logarithmic values of ${{\alpha}_{k}}\left( s \right)$, ${{\beta }_{k}}\left( s \right)$, and ${{\gamma }_{k}}\left( s',s \right)$, respectively. The \emph{a posteriori} LLRs for information bits are computed as
\begin{equation}
\label{Eq:posteriori llr2}
\begin{aligned}
L\left( \left. {{u}_{k}} \right|\mathbf{y} \right)&=\underset{\left( {s}',s \right)\in {{S}^{1}}}{\mathop{{{\max }^{*}}}}\,\left( {{{\bar{\alpha }}}_{k-1}}\left( {{s}'} \right)+{{{\bar{\gamma }}}_{k}}\left( {s}',s \right)+{{{\bar{\beta }}}_{k}}\left( s \right) \right)\\
&-\underset{\left( {s}',s \right)\in {{S}^{0}}}{\mathop{{{\max }^{*}}}}\,\left( {{{\bar{\alpha }}}_{k-1}}\left( {{s}'} \right)+{{{\bar{\gamma }}}_{k}}\left( {s}',s \right)+{{{\bar{\beta }}}_{k}}\left( s \right) \right).
\end{aligned}
\end{equation}

The max-log-MAP algorithm omits the logarithmic term in the Jacobian logarithmic function~\cite{IEEEReduced:Erfanian}. Hence, (\ref{Eq:log_alpha})-(\ref{Eq:posteriori llr2}) can be approximately written as:
\begin{equation}
\label{Eq:max log alpha}
\begin{aligned}
{{\bar{\alpha }}_{k}}\left( s \right)=\underset{{s}'\in {{S}_{R}}}{\mathop{\max }}\,\left( {{{\bar{\alpha }}}_{k-1}}\left( {{s}'} \right)+{{{\bar{\gamma }}}_{k}}\left( {s}',s \right) \right),
\end{aligned}
\end{equation}
\begin{equation}
\label{Eq:max log beta}
\begin{aligned}
{{\bar{\beta }}_{k-1}}\left( {{s}'} \right)=\underset{s\in {{S}_{R}}}{\mathop{\max }}\,\left( {{{\bar{\beta }}}_{k}}\left( s \right)+{{{\bar{\gamma }}}_{k}}\left( {s}',s \right) \right),
\end{aligned}
\end{equation}
and
\begin{equation}
\label{Eq:max_posteriori llr2}
\begin{aligned}
L\left( \left. {{u}_{k}} \right|\mathbf{y} \right)&=\underset{\left( {s}',s \right)\in {{S}^{1}}}{\mathop{\max }}\,\left( {{{\bar{\alpha }}}_{k-1}}\left( {{s}'} \right)+{{{\bar{\gamma }}}_{k}}\left( {s}',s \right)+{{{\bar{\beta }}}_{k}}\left( s \right) \right)\\
&-\underset{\left( {s}',s \right)\in {{S}^{0}}}{\mathop{\max }}\,\left( {{{\bar{\alpha }}}_{k-1}}\left( {{s}'} \right)+{{{\bar{\gamma }}}_{k}}\left( {s}',s \right)+{{{\bar{\beta }}}_{k}}\left( s \right) \right).
\end{aligned}
\end{equation}
In Section~\ref{subsection:dnn turbo decoder}, we provide an alternative graphical representation called neural max-log-MAP algorithm to replace the traditional max-log-MAP algorithm.

\subsection{TurboNet Architecture}
\label{subsection:dnn turbo decoder}
\begin{figure}[htb]
	\centering
	\includegraphics[width=3.5in]{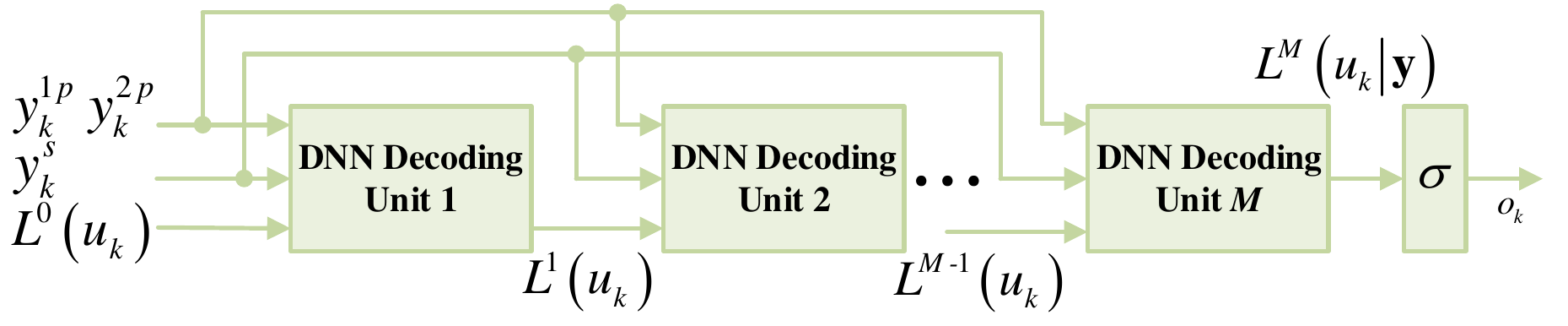}
	\caption{TurboNet architecture. Each DNN decoding unit stands for one iteration. The output of DNN decoding $M$ is ${{L}^{M}}\left( \left. {{u}_{k}} \right|\mathbf{y} \right)$ rather than ${{L}^{M}}\left( {{u}_{k}} \right)$; and \emph{a priori} probability LLRs ${{L}^{0}}\left( {{u}_{k}} \right)$, $k=1,~2,~\ldots,~K$ are initialized to 0.}
	\label{Fig:dnn_turbo_decoder}
\end{figure}

\begin{figure}[htb]
	\centering
	\includegraphics[width=3.5in]{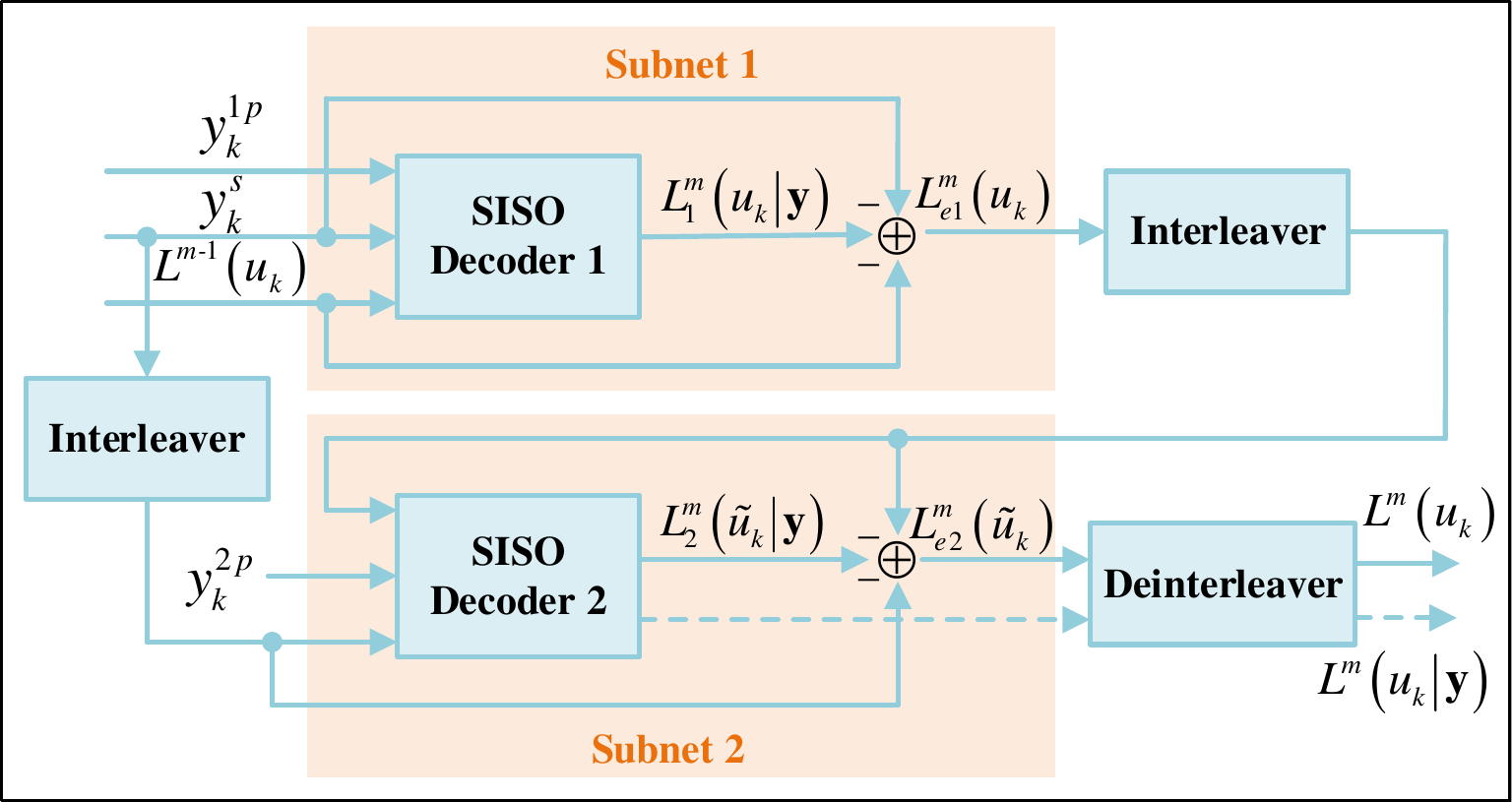}
	\caption{DNN decoding unit \emph{m}. The conventional SISO decoders using max-log-MAP algorithm are replaced by two subnets based on neural max-log-MAP algorithm, as shown in Fig.~\ref{Fig:dnn decoder}.}
	\label{Fig:dnn_turbo_unit}
\end{figure}

The traditional iterative structure is unfolded and each iteration is represented by a DNN decoding unit to obtain an ``unrolled'' structure shown in Fig.~\ref{Fig:dnn_turbo_decoder}, which is equivalent to $M$ iterations. ${{{L}^{m}} ({{u}_{k}})}$ denotes the \emph{a priori} probability LLR calculated by max-log-MAP algorithm with $m$ iterations and ${{{L}^{M}}({{u}_{k}}|\mathbf{y})}$ denotes the \emph{a posteriori} LLR calculated by max-log-MAP algorithm with $M$ iterations, where ${ m=0,~1,~\dots,~M-1}$. The structure of the DNN decoding unit $m$ in Fig.~\ref{Fig:dnn_turbo_decoder} is shown in Fig.~\ref{Fig:dnn_turbo_unit}.

Fig.~\ref{Fig:dnn decoder} shows that subnet 1, which is based on neural max-log-MAP algorithm, consists of $K+3$ layers, $K+1$ of which are hidden layers. Subnet 2 has the same structure as subnet 1. The details of the subnet architecture are elaborated as follows:
\subsubsection{Input Layer} 
The input layer of the proposed network consists of $3K$ neurons, and the output of all neurons constitutes the set ${{{N}^{\rm In}}=\{{{I}_{k}}:k=1,~2,~\ldots,~K\}}$, where ${{{I}_{k}}= \{y_{k}^{s},y_{k}^{1p},L({{u}_{k}})\}}$.
\subsubsection{Hidden Layer 1} 
The first hidden layer consists of $16K$ neurons, and the output of all neurons constitutes the set ${{{N}^{1}}= \{{{{\bar{\gamma}}}_{k}}({s}',s):({s}',s)\in S,k=1,~2,~\ldots ,~K\}}$, where ${S={{S}^{1}}\cup {{S}^{0}}}$ is the set of ordered pairs ${({s}',s)}$ corresponding to all state transitions ${s}'\to s$ caused by data input ${{u}_{k}}$. Some neuron corresponding to ${{{\bar{\gamma}}_{{{k}_{0}}}}({{s}'_{0}},{{s}_{0}})\in {{N}^{1}}}$ in this layer is connected to neurons that corresponding to ${y_{{{k}_{0}}}^{s}}$, ${y_{{{k}_{0}}}^{1p}}$, and ${L({{u}_{{{k}_{0}}}})}$ in the input layer, where ${({{s}'_{0}},{{s}_{0}})\in S}$ and ${{{k}_{0}}\in \{ 1,~2,~\ldots ,~K \}}$.
\subsubsection{Hidden Layer from 2 to K} 
Each layer of the following $K-1$ hidden layers contains 16 neurons. For the $z$th hidden layer, the output of all neurons constitutes the set ${{N}^{z}}=N_{odd}^{z}\cup N_{even}^{z}$, where ${N_{odd}^{z}=\{{{{\bar{\alpha }}}_{k}}(s):k=z-1,s\in{{S}_{R}}\}}$ is the set of neuron outputs for all odd positions in the $z$th hidden layer, ${N_{even}^{z}= \{{{{\bar{\beta}}}_{k-1}}({{s}'}):k=K-z+2,{s}'\in {{S}_{R}}\}}$ is the set of neuron outputs for all even positions in the $z$th hidden layer, and $z=2,~3,~\ldots,~K$. For some ${{{z}_{0}}\in \{2,~3,~\ldots ,~K\}}$, some neuron corresponding to ${{{\bar{\alpha}}_{{{k}_{0}}}}({{s}_{0}})\in N_{odd}^{{{z}_{0}}}}$ in the ${{z}_{0}}$th layer is connected to all neurons corresponding to elements in the set ${\{{{{\bar{\alpha }}}_{{{k}_{0}}-1}} ( {{s}'}):( {s}',{{s}_{0}})\in S \}}$ in layer ${{z}_{0}-1}$ and all neurons corresponding to elements in the set ${\{{{{\bar{\gamma}}}_{{{k}_{0}}}} ( {s}',{{s}_{0}}): ( {s}',{{s}_{0}})\in S \}}$ in the first hidden layer, where ${{{k}_{0}}={{z}_{0}}-1}$ and ${{{s}_{0}}\in {{S}_{R}}}$; some neuron corresponding to ${{{\bar{\beta }}_{{{k}_{1}}-1}}({{s}'_{0}})\in N_{even}^{{{z}_{0}}}}$ in the ${{z}_{0}}$th layer is connected to all neurons corresponding to elements in the set ${\{{{{\bar{\beta }}}_{{{k}_{1}}}}(s):({{s}'_{0}},s)\in S\}}$ in layer ${{z}_{0}-1}$ and all neurons corresponding to elements in the set ${\{{{{\bar{\gamma}}}_{{{k}_{1}}}} ({s}'_{0},s):({s}'_{0},s)\in S \}}$ in the first hidden layer, where ${{{k}_{1}}=K-{{z}_{0}}+2}$ and ${{{s}'_{0}}\in {{S}_{R}}}$. 
\subsubsection{Hidden Layer $K+1$}
The last hidden layer consists of $K$ neurons, and the output of all neurons constitutes the set ${{{N}^{K+1}}=\{L({{u}_{k}}|\mathbf{y}):k=1,~2,~\ldots ,~K\}}$. Some neuron corresponding to ${L({{u}_{{{k}_{0}}}} |\mathbf{y})\in {{N}^{K+1}}}$ in the last hidden layer is connected to all neurons corresponding to elements in the set ${ \{ {{{\bar{\alpha }}}_{{{k}_{0}}-1}} ( {{s}'}):{s}'\in {{S}_{R}} \}}$, $ {\{ {{{\bar{\gamma }}}_{{{k}_{0}}}} ( {s}',s) : ( {s}',s)\in S \}}$, and ${ \{ {{{\bar{\beta }}}_{{{k}_{0}}}} ( s) :s\in {{S}_{R}} \}}$, where ${{{k}_{0}}\in \{1,~2,~\ldots ,~K\}}$. 
\subsubsection{Output Layer}
The output layer consists of $K$ neurons, and the output of all neurons constitutes the set ${{{N}^{\rm Out}}=\{{{L}_{e}}({{u}_{k}}):k=1,~2,~\ldots ,~K\}}$. Some neuron corresponding to ${{{L}_{e}} ( {{u}_{{{k}_{0}}}})\in {{N}^{\rm Out}}}$ is connected to the neuron corresponding to ${L({{u}_{{{k}_{0}}}} |\mathbf{y})}$ in hidden layer $K+1$ and the neurons corresponding to ${y_{{{k}_{0}}}^{s}}$, ${L( {{u}_{{{k}_{0}}}})}$ in the input layer.
\begin{figure}[htb]
	\centering
	\includegraphics[width=2.5in]{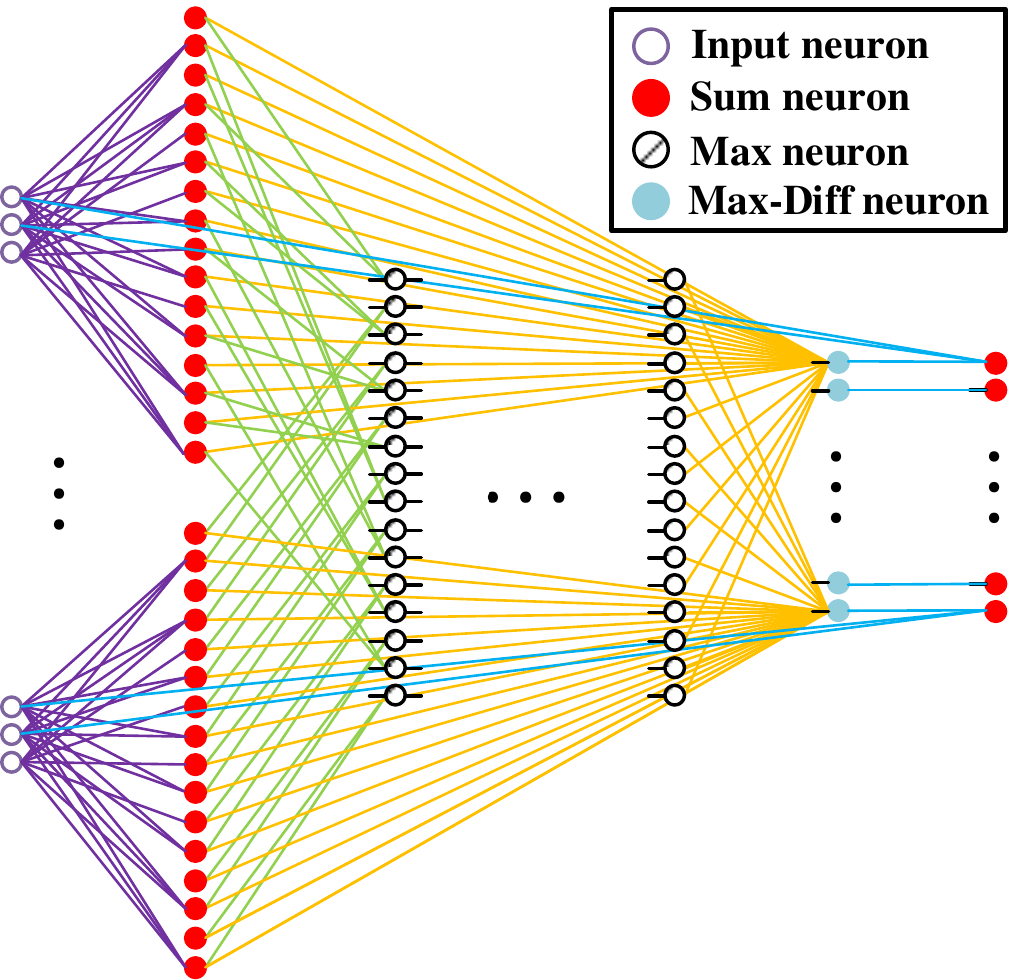}
	\caption{Subnet architecture based on neural max-log-MAP algorithm. The input neuron represents the input value of the network. The sum neuron implements a weighted summation of the input data without an activation function. The max neuron calculates the maximum sum of the input groups. The max-diff neuron realizes that the input data are divided into two categories, and the maximum values of the sum of the groups in both categories are obtained to calculate the difference. The second hidden layer is calculated with initializations ${{{\bar{\alpha}}_{0}}(0)=0}$, ${{{\bar{\alpha}}_{0}}(n)=-128}$ for $n\ne 0$, and ${{{\bar{\beta}}_{K}} (0)=0}$, ${{{\bar{\beta}}_{K}}(n)=-128}$ for $n\ne 0$. Some of the undrawn connecting lines are plotted as small bold lines.}
	\label{Fig:dnn decoder}
\end{figure}

We assign weights to part of the edges in Fig.~\ref{Fig:dnn decoder}. These weights will be trained using the stochastic gradient descent (SGD) algorithm. Therefore, we can calculate ${{{\bar{\gamma }}_{k}}(s',s)}$, ${L({{u}_{k}})}$, and ${{{L}_{e}}( {{u}_{k}})}$ as follows:
\begin{equation}
\label{Eq:weigh gamma}
\begin{aligned}
{{\bar{\gamma }}_{k}}\left( {s}',s \right)=\frac{1}{2}w_{\bar{\gamma },k}^{1}{{u}_{k}}L\left( {{u}_{k}} \right)+\frac{1}{2}w_{\bar{\gamma },k}^{2}x_{k}^{s}y_{k}^{s}+\frac{1}{2}w_{\bar{\gamma },k}^{3}x_{k}^{1p}y_{k}^{1p},
\end{aligned}
\end{equation}
\begin{equation}
\label{Eq:weigh posteriori llr}
\begin{aligned}
L\left( \left. {{u}_{k}} \right|\mathbf{y} \right)&=\hspace{-0.2cm}\underset{\left( {s}',s \right)\in {{S}^{1}}}{\mathop{\max }}\,\hspace{-0.1cm}\left( w_{k}^{1}{{{\bar{\alpha }}}_{k-1}}\left( {{s}'} \right)+w_{k}^{2}{{{\bar{\gamma }}}_{k}}\left( {s}',s \right)+w_{k}^{3}{{{\bar{\beta }}}_{k}}\left( s \right) \right) \\ 
& -\hspace{-0.2cm}\underset{\left( {s}',s \right)\in {{S}^{0}}}{\mathop{\max }}\,\hspace{-0.1cm}\left( w_{k}^{4}{{{\bar{\alpha }}}_{k-1}}\left( {{s}'} \right)+w_{k}^{5}{{{\bar{\gamma }}}_{k}}\left( {s}',s \right)+w_{k}^{6}{{{\bar{\beta }}}_{k}}\left( s \right) \right),   
\end{aligned}
\end{equation}
and
\begin{equation}
\label{Eq:weigh extrinsic llr}
\begin{aligned}
{{L}_{e}}\left( {{u}_{k}} \right)=w_{e,k}^{1}L(\left. {{u}_{k}} \right|\mathbf{y})-w_{e,k}^{2}y_{k}^{s}-w_{e,k}^{3}L({{u}_{k}}).
\end{aligned}
\end{equation}

Turbo code usually has a large block size. For example, the minimum message bit length of Turbo code in the long-term evolution (LTE) standard is 40, and the maximum is 6144. Therefore, parameterizing (\ref{Eq:max log alpha}) and (\ref{Eq:max log beta}) will cause the neural network in Fig.~\ref{Fig:dnn decoder} to be extremely ``deep'', which may lead to gradient vanishing or gradient exploding. On this basis, we do not introduce any trainable parameters.

Given that the output of the $M$th DNN decoding unit is ${{{L}^{M}}({{u}_{k}} |\mathbf{y})}$, the sigmoid function ${\sigma(x)\equiv{{( 1+{{e}^{-x}})}^{-1}}}$ is added, such that the final network output ${{{o}_{k}}=\sigma({{L}^{M}}({{u}_{k}} |\mathbf{y}))}$ is in the range of $[0,~1]$. 
Generally, the mean square error and binary cross-entropy can be used to calculate the network loss with ${{o}_{k}}$ and ${{u}_{k}}$ but for the following two reasons:
\begin{enumerate}
	\item[$\bullet$] The magnitude of the \emph{a posteriori} LLR calculated by the traditional max-log-MAP algorithm is usually greater than 10, whereas the sigmoid function is nearly close to 1 and 0 when $|x|>10$. Therefore, gradient vanishing is likely to occur if the loss is calculated with ${{o}_{k}}$;
	\item[$\bullet$] The loss of the network mainly comes from the occurrence of a few error bits. Hence, the loss of the network becomes extremely small, thereby making the entire network difficult to train.
\end{enumerate}
A redefined loss function computed as (\ref{Eq:loss}) is used to evaluate the loss of TurboNet
\begin{equation}
\label{Eq:loss}
\begin{aligned}
{\rm Loss}=\frac{1}{K}\sum\limits_{k=1}^{K}{{{\left( {{L}^{M}}\left( \left. {{u}_{k}} \right|\mathbf{y} \right)-{{{L}_{\rm log-MAP}} ({{u}_{k}} |\mathbf{y})} \right)}^{2}}},
\end{aligned}
\end{equation}
where ${{L}^{M}}({{u}_{k}}|\mathbf{y})$ represents the \emph{a posteriori} LLR obtained by TurboNet consisting of $M$ decoding units, and ${{{L}_{\rm log-MAP}} ({{u}_{k}} |\mathbf{y})}$ represents the \emph{a posteriori} LLR calculated by the traditional log-MAP algorithm with given iterations.

The goal is to make the loss of the network as small as possible by training the parameters $\{ w_{\bar{\gamma} ,k}^{i},w_{k}^{j},w_{e,k}^{l}\}$. The final decoding results can be obtained by hard decision, as shown as follows:
\begin{equation}
\label{Eq:hard decision}
{{\hat{u}}_{k}}=\left\{ \begin{aligned}
& 1\text{\qquad }{{o}_{k}}\ge 0.5 \\ 
& 0\text{\qquad }{{o}_{k}}<0.5.
\end{aligned} \right.
\end{equation}

By setting all weights to 1, the results of~(\ref{Eq:weigh gamma})-(\ref{Eq:weigh extrinsic llr}) are the same as the original max-log-MAP algorithm. Hence, the performance of the neural max-log-MAP algorithm will not be inferior to the max-log-MAP algorithm by training the network parameters. Moreover, the complexity of TurboNet is similar to the turbo decoder using the max-log-MAP algorithm.

\section{Simulation Results}
\subsection{Parameter Settings}
TurboNet was constructed on top of the TensorFlow framework, and an NVIDIA GeForce GTX 1080 Ti GPU was used for accelerated training. We trained TurboNet for Turbo codes~(40,~132) and~(40,~92) on randomly generated training data obtained over an AWGN channel at 0~dB SNR. TurboNet was composed of three DNN decoding units that corresponded with three full iterations. The loss function~(\ref{Eq:loss}) was used with the target LLR ${{{L}_{\rm log-MAP}} ({{u}_{k}} |\mathbf{y})}$ being the log-MAP algorithm with six iterations. We trained TurboNet with SGD and the ADAM optimizer~\cite{IEEEadam:Kingma} with a batch size of 500. The learning rate was ${{10}^{-5}}$.
\subsection{Performance Analysis}
\begin{figure}[htb]
	\centering
	\includegraphics[width=3in]{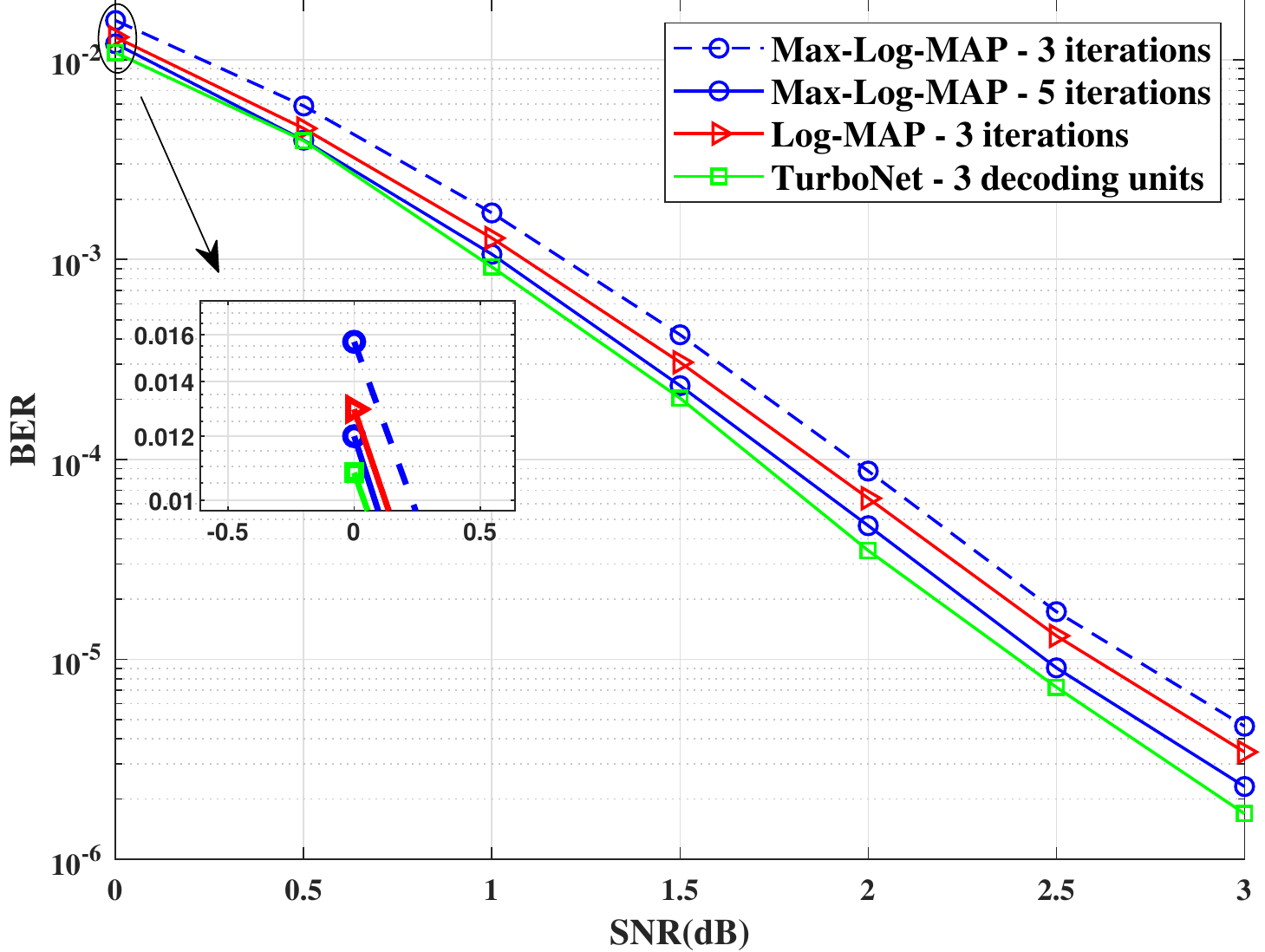} 
	\caption{BER performance curve for Turbo~(40,~132) code using BPSK mapping. The number of training epochs is 50.}
	\label{Fig:ber bpsk}
\end{figure}

\begin{figure}[htb]
	\centering
	\includegraphics[width=3in]{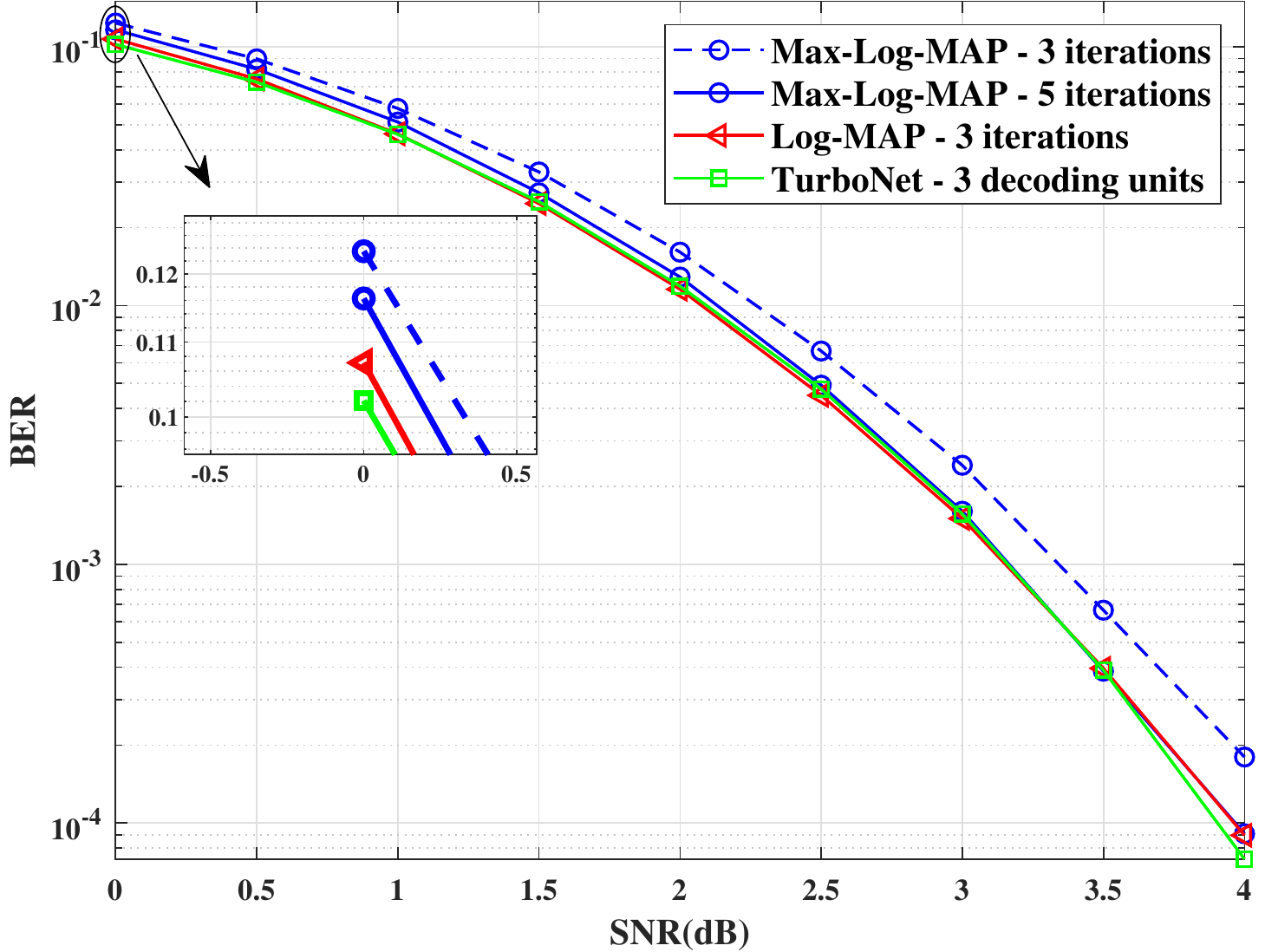}
	\caption{BER performance curve for punctured Turbo~(40,~92) code using BPSK mapping. The number of training epochs is 50.}
	\label{Fig:ber qpsk}
\end{figure}

The bit error rate (BER) performance curves obtained using the log-MAP algorithm, max-log-MAP algorithm, and TurboNet are shown in Figs.~\ref{Fig:ber bpsk} and~\ref{Fig:ber qpsk}.
\subsubsection{BER Performance}
Fig.~\ref{Fig:ber bpsk} indicates that the BER of TurboNet that consists of three decoding units is lower than those of the max-log-MAP algorithm and the log-MAP algorithm with three iterations at all SNR ranges. Notably, TurboNet also outperforms the max-log-MAP algorithm with five iterations in almost all cases. Fig.~\ref{Fig:ber qpsk} shows that TurboNet containing three decoding units outperforms the max-log-MAP algorithm with three iterations at all SNR ranges. The BER performance of TurboNet is comparable to that of the log-MAP algorithm with three iterations and the max-log-MAP algorithm with five iterations under most circumstances. These results suggest that TurboNet can still work when handling punctured Turbo code with a high code rate.

Here, we detail about the SNR of the training data and the iteration number of the log-MAP algorithm in~(\ref{Eq:loss}), which are closely related to the training of TurboNet.
\begin{enumerate}
	\item[$\bullet$] The training data and the test data can have a similar distribution; thus, one might use the same SNR for testing and training, which is restricted because the precise SNR might not be available. Moreover, TurboNet is equivalent to traditional max-log-MAP algorithm by setting all weights to 1. Therefore, TurboNet is not able to learn to handle noise when the SNR is too high. However, if the SNR is set too low, the max-log-MAP algorithm has poor error correction capabilities and thus cannot learn effectively. So we deliberately reduced the SNR such that TurboNet could learn more robust error correction. Hence, we keep the SNR of the training data at a single 0 dB, which can help TurboNet learn to correct errors as much as possible.
	\item[$\bullet$] Notably, the target LLR values in~(\ref{Eq:loss}) are generated by the log-MAP algorithm with a fixed iteration number $T$. If the iteration number is large, then TurboNet can learn accurate information. However, $T$ should not be too large because TurboNet only contains three decoding units. If $T$ is too large, then a large gap will exist between TurboNet and the log-MAP algorithm, thereby decreasing the generalization capability of TurboNet. Therefore, we set $T=6$, which is exactly twice the number of the decoding units.
\end{enumerate}

The improvement of BER is achieved by properly configuring the weight, such that the logarithmic term in the Jacobian logarithmic function is compensated appropriately. In addition, the LLRs $\{ y_{k}^{s},y_{k}^{1p},y_{k}^{2p}\}$ are related to the channel conditions; thus, the part of the channel information might be learned by TurboNet, thereby making these LLRs used precisely.
\begin{table}[htb]
	\caption{Complexity analysis for Turbo decoder}
	\label{tab:completixexity}
	\centering
	\begin{tabular}{ccc}
		\toprule
		& \# of parameters & Time\\
		\hline
		Max-Log-MAP (5 iterations) & - & 2.3e-4s\\
		Neural BCJR in\cite{IEEEcommunicationalgorithms:Kim} (3 units) & 3.85M & 5.89e-3s\\
		TurboNet (3 decoding units) & 17.8K & 1.39e-4s\\
		\bottomrule
	\end{tabular}
\end{table}
\subsubsection{Computational Complexity}
Table~\ref{tab:completixexity} compares the complexities of the decoders in terms of the number of parameters and time consumption required to complete a single-forward pass of one codeword. TurboNet has a lower computational cost and exhibits relatively faster computation speed with considerably fewer parameters compared with the data-driven neural BCJR decoder~\cite{IEEEcommunicationalgorithms:Kim}. The SISO decoder in the neural BCJR decoder is replaced by two bidirectional LSTM layers, and the number of hidden units in each LSTM layer is 800. In addition, TurboNet shows lower latency compared with the max-log-MAP algorithm with five iterations.
\section{Conclusion}
In this work, we demonstrate the benefits of the proposed
TurboNet decoder architecture compared with traditional turbo decoder based on the max-log-MAP algorithm. In TurboNet, the original iterative structure is unfolded, and each iteration is represented as a DNN decoding unit. We obtain a neural max-log-MAP algorithm by assigning weights to the max-log-MAP algorithm. Moreover, a redefined loss function is used to improve the training process. The BER performance of TurboNet is improved compared with the max-log-MAP algorithm without increasing computational complexity. The error correction capability of the proposed TurboNet can be further improved by applying advanced DL technology, and we hope this letter encourages future research in this direction.


\begin{thebibliography}{1}
	
	
	\bibitem{IEEEondeep:Gruber}
	T.~Gruber, S.~Cammerer, J.~Hoydis, and S.~ten Brink, ``On deep learning-based channel decoding,'' in \emph{Proc. IEEE 51st Annu. Conf. Inf. Sciences Syst.}, Mar.~2017, pp.~1-6.
	
	\bibitem{IEEEanartificial:Wang}
	X.-A.~Wang and S.~B.~Wicker, ``An artificial neural net Viterbi decoder,'' \emph{IEEE Trans. Commun.}, vol.~44, no.~2, pp.~165-171, Feb.~1996.
	
	\bibitem{IEEEcommunicationalgorithms:Kim}
	H.~Kim, Y.~Jiang, R.~B.~Rana, S.~Kannan, S.~Oh, and P.~Viswanath, ``Communication algorithms via deep learning,'' \emph{arxiv preprint arXiv:1805.09317},~2018. 
	
	\bibitem{IEEElearningto:Nachmani}
	E.~Nachmani, Y.~Be’ery, and D.~Burshtein, ``Learning to decode linear codes using deep learning,'' in \emph{Proc.~IEEE Annu. Allerton Conf. Commun., Control, and Computing}, 2016, pp.~341-346.
	
	\bibitem{IEEEneuraloffset:Lugosch}
	L.~Lugosch and W.~J. Gross, ``Neural offset min-sum decoding,'' in \emph{Proc. 2017 IEEE Int. Symp. Inf. Theory}, Jun.~2017, pp.~1361-1365.
	
	\bibitem{IEEEdeeplearning:Nachmani}
	E.~Nachmani, E.~Marciano, L.~Lugosch, W.~J. Gross, D.~Burshtein, and Y.~Be’ery, ``Deep learning methods for improved decoding of linear codes'', \emph{IEEE J. Sel. Topics Signal Process}., vol.~12, no.~1, pp.~119-131, Feb.~2018.
	
	\bibitem{IEEE3GPP}
	3rd Generation Partnership Project; Technical Specification; Evolved Universal Terrestrial Radio Access (E-UTRA); Multiplexing and Channel Coding (Release 9) 3GPP Organizational Partners TS 36.212, Rev. 8.3.0, May 2008. 
	
	\bibitem{IEEEturbocode:Berrou}
	C.~Berrou, A.~Glavieux, and P.~Thitimajshima, ``Near Shannon limit error-correcting coding and decoding: Turbo-codes,'' in \emph{Proc.~Int. Conf. Communications}, May~1993, pp.~1064-1070.
	
	\bibitem{IEEEoptimaldecoding:Bahl}
	L.~Bahl, J.~Cocke, F.~Jelinek, and J.~Raviv, ``Optimal decoding of linear codes for minimizing symbol error rate,'' \emph{IEEE Trans. Inf. Theory}, vol.~IT-20, no.~2, pp.~284-287, Mar.~1974.
	
	\bibitem{IEEEiterative:Bauch}
	S.~Talakoub, L.~Sabeti, B.~Shahrrava, and M.~Ahmadi, ``An improved Max-Log-MAP algorithm for turbo decoding and turbo equalization,'' \emph{IEEE Trans. Instrum. Meas.}, vol.~56, no.~3, pp.~1058-1063, June 2007.
	
	\bibitem{IEEEOptimal:Robertson}
	P.~Robertson, P.~Hoeher, and E.~Villebrun, ``Optimal and sub-optimal maximum a posteriori algorithms suitable for turbo decoding,'' \emph{Eur. Trans. Telecommun.}, vol.~8, no.~2, pp.~119-125, 1997.
	
	\bibitem{IEEEReduced:Erfanian}
	J.~A.~Erfanian, S.~Pasupathy, and G.~Gulak, ``Reduced complexity symbol detectors with parallel structures for ISI channels,'' \emph{IEEE Trans. Commun.}, vol.~42, no.~2-4, pp.~1661-1671, Feb.-Apr.~1994.
	
	
	\bibitem{IEEEadam:Kingma}
	D.~P.~Kingma and J.~Ba. (2014). ``Adam: A method for stochastic optimization.'' \emph{arxiv preprint arXiv:1412.6980},~2014. 
\end{thebibliography}
\end{document}